\documentclass[11pt,a4paper]{article}
\usepackage{jheppub,xcolor}
\usepackage[utf8]{inputenc}
\usepackage{textcomp}
\usepackage{newunicodechar}
\newunicodechar{γ}{\ensuremath{\gamma}}
\usepackage{blindtext}
\usepackage{graphicx}
\usepackage{physics}
\usepackage{amsmath}
\usepackage{amsfonts}
\usepackage{amsthm}
\usepackage{amssymb}
\usepackage{caption}
\usepackage{booktabs}
\usepackage{subcaption}
\usepackage{slashed}
\usepackage{bigints}
\usepackage{setspace}
\usepackage{array}
\usepackage{comment}
\usepackage{multirow}
\usepackage{makecell} 
\definecolor{mygreen}{rgb}{0, 0.5019607843137255, 0}
\definecolor{Red}{rgb}{1.,0.,0.}
\definecolor{Blue}{rgb}{0.,0.,1.}
\newcommand{\CF}{C_F}
\newcommand{\CA}{C_A}
\newcommand{\qbar}{\overline{Q}}

\newcommand{\tbar}{\Bar{t}}
\newcommand{\bbar}{\Bar{b}}

\def\beq{\begin{equation}}
\def\eeq{\end{equation}}
\def\bsp#1\esp{\begin{split}#1\end{split}}
\newcommand{\rd}{\textrm{d}}
\newcommand{\ord}{\mathcal{O}}
\newcommand{\cO}{\mathcal{O}}
\newcommand{\MSbar}{$\overline{\textrm{MS}}$}

\allowdisplaybreaks

\title{Two-loop renormalisation of quark and gluon fields in the SMEFT in the
on-shell scheme}

\author[a]{Claude Duhr,}
\author[b]{Giuseppe Ventura,}
\author[b]{Eleni Vryonidou}

\emailAdd{cduhr@uni-bonn.de} \emailAdd{giuseppe.ventura@manchester.ac.uk} \emailAdd{eleni.vryonidou@manchester.ac.uk}
\affiliation[a]{Bethe Center for Theoretical Physics, Universit\"{a}t Bonn, Wegelerstrasse 10, D-53115, Germany}
\affiliation[b]{Department of Physics and Astronomy, University of Manchester, Oxford Road, Manchester M13~9PL, United Kingdom}

\abstract{ We compute the contributions of CP-conserving dimension-six SMEFT operators to the two-loop renormalisation constants of quark and gluon fields in the on-shell
 scheme. Specifically, we consider the top-quark chromomagnetic operator and the triple gluon operator. We also compute the contribution of four-quark operators to the gluon renormalisation constant and discuss the implications for the running of the strong coupling constant. 

}

\begin{document}

\begin{flushright}
BONN-TH-2025-27
\end{flushright}

\maketitle

\section{Introduction}
Collider experiments scrutinising the predictions of the Standard Model (SM) and searching for new particles have not found any evidence for new light degrees of freedom. This has motivated the use of the Standard Model Effective Field Theory (SMEFT) framework \cite{Weinberg:1979sa, Leung:1984ni, Buchmuller:1985jz, 1008.4884} to parametrise new physics effects through higher-dimensional operators modifying the interactions of the SM particles in a model-independent manner.  

Significant effort has been devoted to both interpreting experimental data within the context of the SMEFT in global fits \cite{2012.02779,Ethier:2021bye, Celada:2024mcf, 2507.06191}, and to improving theoretical predictions to maximise our sensitivity to new physics through these interpretations. A campaign of higher-order SMEFT computations have appeared in the literature, with one-loop corrections in the QCD coupling automated with \texttt{SMEFT@NLO} \cite{2008.11743}. In parallel, a significant endeavour to extract the anomalous dimension matrix for the dimension-six Wilson coefficients at higher loop orders is currently on-going. Whilst this is known for the SMEFT at one-loop \cite{1308.2627, 1310.4838, 1312.2014}, promoting these results to the two-loop order is a big undertaking currently addressed by different approaches \cite{1910.05831,2005.12917,2308.06315, 2310.19883, 2311.13630,2211.09144, 2203.11224,2401.16904, 2408.03252, 2410.07320,2412.13251, 1907.04923,2011.02494,Fuentes-Martin:2024agf, 2501.08384, 2502.14030, Haisch:2025lvd,2504.18537,2504.00792,Naterop:2025lzc,2507.08926,2507.10295,2507.19589, 2507.12518,Haisch:2025vqj}. 

A related and important ingredient towards the computation of two-loop corrections in the SMEFT is the renormalisation of the SM fields and masses, as well as that of the strong coupling constant, including the effects of higher-dimension operators. Explicit results in this direction have appeared in the literature both in the context of LEFT \cite{2412.13251,2507.08926} and SMEFT \cite{Duhr:2025zqw,2507.10295}. In particular, in a recent paper~\cite{Duhr:2025zqw} we have computed the two-loop QCD corrections to the quark and gluon self-energies, including the contribution from a specific subset of SMEFT dimension-six operators in the modified minimal-subtraction ($\overline{\text{MS}}$) scheme. The operators considered in ref.~\cite{Duhr:2025zqw} were the top quark chromomagnetic dipole moment and the triple-gluon operator. Our computation employed the background field method, which also allowed us to determine the contribution of these operators to the QCD $\beta$-function at two loops. 

In this work we extend our previous work in two important directions. First, we perform the computation of the field renormalisation constants in the on-shell scheme. In particular, we extract the renormalisation constants for the quark and gluon fields within the on-shell scheme and the relation between the pole and the $\overline{\text{MS}}$ top masses. These results complement our previous work and allow flexibility in their potential applications in the context of computations of cross-sections for physical processes such as top pair production, for which the SM computations employ the on-shell scheme, see for example ref.~\cite{0707.4139}. 

The second direction we pursue is the extension of our operator basis to include four-quark operators. The presence of chirality-dependent operators introduces a significant complication into the calculation, due to the treatment of the $\gamma_5$ matrix beyond four dimensions. Several schemes have been proposed in the literature to address this issue, with the Breitenlohner-Maison-'t Hooft-Veltman (BMHV) scheme~\cite{tHooft:1972tcz, Breitenlohner:1977hr} being the only scheme shown to be algebraically consistent at any loop order. In addition to the treatment of $\gamma_5$ in dimensional regularisation, special care is needed as the definition of a basis of operators is based on the use of spinor identities which only hold in four dimensions. As such, one is forced to also consider evanescent operators in the computation. Both aspects significantly increase the computational complexity of this two-loop renormalisation programme. In this work we focus on the renormalisation of the gluon field in the BMHV scheme, addressing all issues arising in the presence of the four-fermion operators. We also use our results to comment on the running of the strong coupling constant in the presence of dimension-six operators. In an appendix we also discuss how, by appropriately defining an effective coupling, we can decouple the effects of dimension-six contributions from the running of the strong coupling constant. 

This work is organised as follows. We start by introducing the operators we consider in this paper and our computational setup in section \ref{sec:SMEFT_QCD}. In section \ref{sec:on-shell} we extract the calculation of the two-loop renormalisation of gluon and quark fields in the on-shell scheme, and present the relationship between the pole and $\overline{\text{MS}}$ masses for the top quark. Then in section \ref{sec:4F} we discuss the gluon renormalisation in the presence four-quark operators showing results in both the $\overline{\text{MS}}$ and on-shell schemes. We conclude in section \ref{sec: conclusions}. In our appendices we provide additional information about our computation and we discuss also the decoupling scheme for the strong coupling.


\section{The QCD sector of the SMEFT}
\label{sec:SMEFT_QCD}

This paper is part of a sequence of works aiming at the two-loop renormalisation of the QCD sector of the SMEFT with dimension-six operators added. More specifically, we consider the theory defined by the Lagrangian collecting the CP-conserving dimension-six operators of the SMEFT that only involve quark and gluon fields,
\beq\label{eq:LQCD6}
\mathcal{L}_{\textrm{QCD},6}= \mathcal{L}_{\text{QCD}}+\frac{c_{G}^0}{\Lambda^2}\,\mathcal{O}_G+\sum_{i,j=1}^{n_u}\frac{c_{uG}^{0,ij}}{\Lambda^2}\mathcal{O}_{uG}^{ij}+\sum_{i,j=1}^{n_d}\frac{c_{dG}^{0,ij}}{\Lambda^2}\mathcal{O}_{dG}^{ij} + \sum_n\frac{c_{4q_n}^{0}}{\Lambda^2}\mathcal{O}_{4q_n}\,,
\eeq 
where $\mathcal{L}_{\text{QCD}}$ is the SM QCD Lagrangian with $n_f=n_u+n_d=6$ different quark flavours ($n_u=3$ and $n_d=3$ are the number of up and down-type quarks, respectively), $\Lambda$ is the SMEFT scale, and $c_{G}^0$, $c_{uG}^{0,ij}$, $c_{dG}^{0,ij}$ and $c_{4q_n}^0$ are the bare Wilson coefficients. This Lagrangian defines the appropriate setting to study the impact of the SMEFT on the renormalisation of quark and gluon fields and masses, as well as of the strong coupling constant. We now describe the dimension-six operators we consider in eq.~\eqref{eq:LQCD6}.

The operator $\mathcal{O}_G$ is the unique CP-even operator of dimension six that only involves gluon fields,
\beq
\mathcal{O}_G = f^{abc}\,G_{\mu}^{0a\nu}\,G_{\nu}^{0b\rho}\,G_{\rho}^{0c\mu}\,,
\eeq
where $f^{abc}$ are the SU$(N)$ structure constants and $G_{\mu\nu}^{0a}$ is the (bare) gluon field strength tensor,
\beq
G_{\mu\nu}^{0a} = \partial_{\mu}G_{\nu}^{0a}-\partial_{\nu}G_{\mu}^{0a}-g_s^0\,f^{abc}\,G_{\mu}^{0b}\,G_{\nu}^{0c}\,,
\eeq
with $G_{\mu}^{0a}$ and $g^0_s$ the bare gluon field and bare strong coupling, respectively.

The chromomagnetic operators in the SMEFT are defined as
\beq\bsp\label{eq:chromo_def_6}
\widetilde{\mathcal{O}}_{uG}^{ij} &\,= {i}\,\overline{Q}^0_iT^a\tau^{\mu\nu}\widetilde{\phi}\,u^0_{Rj}\,G_{\mu\nu}^{0a} + \textrm{h.c.}\,,\qquad 1\le i,j\le n_u\,,\\
\widetilde{\mathcal{O}}_{dG}^{ij} &\,= {i}\,\overline{Q}^0_iT^a\tau^{\mu\nu}{\phi}\,d^0_{Rj}\,G_{\mu\nu}^{0a} + \textrm{h.c.}\,,\qquad 1\le i,j\le n_d\,,
\esp\eeq
where $Q^0_i$ are the bare left-handed quark doublets, $u^0_{Ri}$ and $d^0_{Ri}$ are the bare right-handed quark singlets and  $\phi$ is the SM Higgs doublet, with $\widetilde{\phi} = i\sigma^2\phi^*$. Moreover, $\tau^{\mu\nu} = \frac{1}{2}[\gamma^{\mu},\gamma^{\nu}]$ and $T^a$ are the generators of the fundamental representation SU$(N)$, normalised according to $\Tr(T^aT^b) = \frac{1}{2}\delta^{ab}$. After electroweak symmetry breaking, these operators lead to the following dimension-five operators,
\beq\bsp\label{eq:chromo_def}
\mathcal{O}_{uG}^{ij} &\,= \frac{iv}{\sqrt{2}}\,\overline{u}^0_iT^a\tau^{\mu\nu}u^0_j\,G_{\mu\nu}^{0a}\,,\qquad 1\le i,j\le n_u\,,\\
\mathcal{O}_{dG}^{ij} &\,= \frac{iv}{\sqrt{2}}\,\overline{d}^0_iT^a\tau^{\mu\nu}d^0_j\,G_{\mu\nu}^{0a}\,,\qquad 1\le i,j\le n_d\,,
\esp\eeq
where $v$ is the vacuum expectation value (vev) of the SM Higgs field, and $u_i^0$ and $d_i^0$ are the bare up- and down-type bare quark fields, respectively.

 Whilst chromomagnetic dipole operators exist for all flavours at dimension six, typically a flavour symmetry is employed which restricts the number of operators considered in any SMEFT analysis. A popular choice is a $U(2)_q\times U(2)_u \times U(2)_d$ flavour symmetry, motivated by a Minimal Flavour Violation ansatz, which excludes chirality flipping bilinears in the first two generations, and thus forbids the flavour diagonal dipole operators of eq.~\eqref{eq:chromo_def} for $i,j\ne 3$ ~\cite{1802.07237}. 

 Finally, the operators $\mathcal{O}_{4q_n}$ involve four quark fields. The  flavour assumption restricts the number of four-fermion operators allowed.  From a phenomenological standpoint, however, not all four-fermion operators allowed by the flavour assumption are equally relevant. In particular, once we consider current constraints from experimental measurements, four-fermion operators with four heavy quark fields are the least constrained ones.  Thus, they are  particularly interesting from a phenomenological point of view. Motivated by this argument, and also to have a manageable number of operators, we will focus on the following class of four-quark operators involving only third-generation quarks:
 \begin{align}
& 
\left.
\begin{aligned}
  \cO_{QQ}^{(1)} &= \frac{1}{2}(\qbar \gamma^\mu Q)(\qbar \gamma_\mu Q)\,, \\
  \cO_{QQ}^{(8)} &= \frac{1}{2}(\qbar \gamma^\mu T^A Q)(\qbar \gamma_\mu T^A Q)\,,
\end{aligned}
\, \,\right\} (\overline{L}L)(\overline{L}L) \notag \\
& 
\left.
\begin{aligned}
  \cO_{Qt}^{(1)} &= (\qbar \gamma^\mu Q)(\tbar_R \gamma_\mu  t_R)\,, \\
  \cO_{Qt}^{(8)} &= (\qbar \gamma^\mu T^A Q)(\tbar_R \gamma_\mu T^A t_R)\,, \\
  \cO_{Qb}^{(1)} &= (\qbar \gamma^\mu Q)(\bbar_R \gamma_\mu  b_R)\,, \\
  \cO_{Qb}^{(8)} &= (\qbar \gamma^\mu T^A Q)(\bbar_R \gamma_\mu T^A b_R)\,,
\end{aligned}
\, \,\right\} (\overline{L}L)(\overline{R}R) \notag \\
&
\left.
\begin{aligned}
  \cO_{tt} &= (\tbar_R \gamma^\mu t_R)(\tbar_R \gamma_\mu t_R)\,, \\
  \cO_{bb} &= (\bbar_R \gamma^\mu b_R)(\bbar_R \gamma_\mu b_R)\,, \\
  \cO_{tb}^{(1)} &= (\tbar_R \gamma^\mu t_R)(\bbar_R \gamma_\mu b_R)\,, \\
  \cO_{tb}^{(8)} &= (\tbar_R \gamma^\mu T^A t_R)(\bbar_R \gamma_\mu T^A b_R)\,,
\end{aligned}
\, \,\right\} (\overline{R}R)(\overline{R}R) \notag \\
&
\left.
\begin{aligned}
  \cO_{QtQb}^{(1)} &= (\qbar^I  t_R)\epsilon_{IJ}(\qbar^J  b_R)\, + \mathrm{h.c.} \,, \\
  \cO_{QtQb}^{(8)} &= (\qbar^I   T^A t_R)\epsilon_{IJ}(\qbar^J T^A b_R)\, + \mathrm{h.c.} \,,
\end{aligned}
\, \,\right\} (\overline{L}R)(\overline{L}R)
\label{eq: four-quark}
\end{align}
where we have introduced the shorthand notation $\{Q,t_R,b_R\}$\footnote{To simplify the notation in eq.~\eqref{eq: four-quark}, we omit the superscript $^0$ denoting bare quantities.} for the bare SM quark fields of the third generation $\{Q^0_3, u^0_{R3},d^0_{R3}\}$, respectively, and $\epsilon_{IJ}$ is the Levi-Civita tensor with two indices.
This choice of basis in eq.~\eqref{eq: four-quark} differs from the Warsaw basis~\cite{1008.4884} by the choice of the $(\overline{L}L)(\overline{L}L)$ operators. In particular, for the colour-singlet operator we have $\cO_{QQ}^{(1)}=\frac{1}{2}( \cO_{qq}^{(1),{\textrm{Warsaw}}})_{3333}$, while the colour octet operator $\cO_{QQ}^{(8)}$ replaces the triplet operator $\cO_{qq}^{(3),{\text{Warsaw}}}$. The choice of basis in eq.~\eqref{eq: four-quark} reproduces the conventions of ref.~\cite{1802.07237} at leading order. We underline that beyond tree-level the definition of the basis becomes subtle, and for example, defining the octet operator as the octet structure itself, or as a linear combination of a singlet and a triplet operators, will lead to different results, because they differ by an evanescent operator. We will further discuss these subtleties in section \ref{sec: gluon renormalisation}, where we also comment on the impact of the rest of the four-fermion operators on the quantities we compute.

As we mentioned in the Introduction, the goal of this paper is to make progress towards our two-loop SMEFT renormalisation programme of the Lagrangian in eq. \eqref{eq:LQCD6}, first by computing the contributions of the operators $\mathcal{O}_{G}$ and $\mathcal{O}_{tG}$ to the renormalisation constants for the fields and the top mass in the on-shell scheme and second, by computing the contributions of the four-fermion operators of eq.~\eqref{eq: four-quark} to the gluon field renormalisation, assuming that both the top and bottom quarks are massive. Before we discuss the computation of these renormalisation constants, it is worth discussing the interplay between the operators we consider and our quark mass and flavour assumptions.
 
At high-energy colliders like the LHC, it is customary to consider all quarks except for the top quark, and possibly the bottom quark, massless. If the bottom quark is considered massless, then it was argued in ref.~\cite{Duhr:2025zqw} that, since only the interference of dimension-six operators and the SM is relevant at this order in the EFT expansion and since the only sources of chirality flips in the SM are mass terms, only the chromomagnetic operator $\mathcal{O}_{tG} = \mathcal{O}_{uG}^{33}$ is relevant. This allows one to neglect all other chromomagnetic operators.

To assess the interplay with the four-fermion operators, we consider the one-loop renormalisation of bottom and top masses in the presence of the  operators of eq. \eqref{eq: four-quark}. To do so, we define 
the renormalisation factors
\begin{equation}
    m_{i}^0 = (Z_{m})_{ij} \, m_{_j}\,, \ \ i \in \{t, b\}\,,
    \label{eq:Z_m_matrix_def}
\end{equation}
where $t$ and $b$ now refer to the physical top and bottom fields, as we work in the broken phase, and $m_{t(b)}$ is the $\overline{\text{MS}}$ renormalised top (bottom) mass, while $Z_{m}$ is a $2 \cross 2$ matrix.  The mass-counterterm matrix can be expanded as
\begin{equation}
    Z_{m} = 1 + \frac{1}{16 \pi^2 }\,\delta Z_{m,\textrm{QCD}}
    + \frac{1}{16 \pi^2 \,\Lambda^2}\,\left[\delta Z_{m,G}
    + \delta Z_{m,qG}
    +\delta Z_{m,4q}\right]+\ord\left(\frac{1}{\Lambda^4}\right)\,,
\end{equation}
where $\delta Z_{m,\textrm{QCD}}$ is the pure QCD contribution and $\delta Z_{m,G}$, $\delta Z_{m,qG}$ and $\delta Z_{m,4q}$ denote the contributions from the triple-gluon, chromomagnetic and four-quark operators, respectively.
With the order $(t, b)$, the one-loop contribution $\delta Z_{m,4q}^{(1)}$ to the four-quark mass counterterm reads
\begin{equation}
\renewcommand{\arraystretch}{2}
\setlength{\arraycolsep}{7pt}
\delta Z_{m,4q}^{(1)}=\frac{1}{\varepsilon}
\begin{pmatrix}
  -4 m_t^2\qty(c_{Qt}^{(1)}+\CF\, c_{Qt}^{(8)}) & m_b^2\left((2 \CA+1)\, c_{QtQb}^{(1)}+\CF\, c_{QtQb}^{(8)}\right) \\ m_t^2\left((2 \CA+1)\, c_{QtQb}^{(1)}+\CF \,c_{QtQb}^{(8)}\right) & -4 m_b^2\qty(c_{Qb}^{(1)}+\CF\, c_{Qb}^{(8)}) 
\end{pmatrix},
\label{eq: mass mixing matrix}
\end{equation}
where the Wilson coefficients $c_{QtQb}^{(R)}$ with $R\in\{1,8\}$ are real as we are interested in the CP-conserving Lagrangian, and we introduced the usual values of the quadratic Casimir operators of the fundamental and adjoint representations of SU$(N)$,
\beq
\CF = \frac{N^2-1}{2N}\textrm{~~~and~~~} \CA = N\,.
\eeq
We note that the field renormalisation factors do not receive any corrections at one loop from the four-fermion operators. In our mass mixing matrix from eq.~\eqref{eq: mass mixing matrix}, the off-diagonal terms are non-zero, and consequently, the non-zero mass of the top quark may radiatively  generate a loop-suppressed mass term for the bottom quark, and vice versa. 

Let us conclude our discussion of the interplay between four-fermion operators and quark masses by making a comment about the connection to QCD factorisation. 
Since a non-zero value for the bottom mass may be generated radiatively, special care is needed when attempting to perform a computation involving the chirality flipping four-fermion operators $\cO^{(R)}_{QtQb}$ in the five flavour scheme (5FS). The 5FS assumes a massless bottom quark and a massive top quark, and consequently includes a bottom quark parton distribution function into the QCD factorisation formula. Since the non-zero top quark mass may radiatively generate a non-zero bottom mass via eq.~\eqref{eq: mass mixing matrix}, we conclude that higher-order computations in the SMEFT using the 5FS are only consistent if the Wilson coefficients are such that the $2\times2$ mass matrix has a zero eigenvalue.

\section{On-shell renormalisation of quark and gluon fields : \texorpdfstring{$\mathcal{O}_{tG}$ and $\mathcal{O}_G$}{OtG and OG}}
\label{sec:on-shell}

In this section we discuss the renormalisation of quark and gluon fields in the on-shell scheme. We work in the five-flavour scheme, and we consider all quarks massless except for the top quark. We also only focus on the triple-gluon and chromomagnetic operators $\cO_G$ and $\cO_{tG}$, similar to our previous results in the $\overline{\textrm{MS}}$ scheme from ref.~\cite{Duhr:2025zqw}. The contributions from the four-fermion operators to the quark-field renormalisation in the on-shell scheme are more complicated, due to the appearance of evanescent operators and symmetry restoring counterterms, and they will be discussed in a dedicated publication~\cite{Duhr:2026btw}. We will discuss the contribution of four-fermion operators to the on-shell renormalisation of the gluon field in section~\ref{sec:4F}.

\subsection{On-shell renormalisation of the top quark}
\label{eq:top_OS}
In this section we discuss the renormalisation of the top quark mass and field in the on-shell scheme, focusing on the effects of the operators $\mathcal{O}_G$ and $\mathcal{O}_{tG}$ defined in section~\ref{sec:SMEFT_QCD}. Due to their structures in the CP-conserving scenario, these operators do not distinguish between the left- and right-handed components of the spinor fields, meaning that we have just one independent field renormalisation factor $Z_{t}$, which relates the bare and renormalised top quark fields,
\beq
t^0 = \sqrt{Z_t}\,t\,.
\eeq

We can write the correction to the top self-energy as
\begin{equation}
    \Sigma(p, m^0) =  \Sigma_{V}(p, m^0) \, \slashed{p} + \Sigma_{S}(p, m^0) \, m^0\,,
    \label{eq: top self-energy-generic}
\end{equation}
where $m^0$ is the bare mass of the top quark. Equation~\eqref{eq: top self-energy-generic} can be rearranged as
\begin{equation}
    \Sigma(p, Z_M M) = \Sigma_1 (p, Z_M M) \, M + \Sigma_2 (p, Z_M M)\, (\slashed{p}-M)\,, 
    \label{eq: top renormalised self-energy-generic}
\end{equation}
where we introduced the renormalised top mass $M$ in the on-shell scheme 
\beq
m^0 = Z_M M\,.
\label{eq:Z_M_def}
\eeq
Comparing eqs.~\eqref{eq: top self-energy-generic} and~\eqref{eq: top renormalised self-energy-generic}, we see that
\begin{equation}
\begin{split}
    & \Sigma_1 (p, Z_M M) = \Sigma_{V}(p, Z_M M) + Z_M \Sigma_{S}(p, Z_M M)\,, \\ &
     \Sigma_2 (p, Z_M M) = \Sigma_{V}(p, Z_M M)\,.
     \label{eq:Sigma1_2}
\end{split}
\end{equation}
As is well known, we can extract the renormalisation conditions in the on-shell scheme by imposing that the renormalised propagator has a single pole with unit residue at the on-shell mass $p^2=M^2$. These read
\begin{equation}
\begin{split}
  &  Z_M = 1 + \eval{\Sigma_1(p, Z_M M)}_{p^2=M^2}, \\
  & \frac{1}{Z_t} = 1+ \eval{2p^2 \frac{\partial}{\partial p^2} \Sigma_1(p, Z_M M)}_{p^2=M^2}+\eval{\Sigma_2(p, Z_M M)}_{p^2=M^2}.
\end{split}
\label{eq: OS renormalisation conditions}
\end{equation}

Before we proceed to describe the computation and present our results, let us briefly compare the computation in the on-shell scheme performed here with the corresponding computation in the \MSbar-scheme of ref.~\cite{Duhr:2025zqw}. At first glance, it appears that the only difference between the computations of the renormalisation constants in the two schemes  lies in the choice of the renormalisation conditions. In particular, the renormalisation conditions in the on-shell scheme from eq.~\eqref{eq: OS renormalisation conditions} are harder to implement than those in the \MSbar-scheme, where only the pole parts of the self-energies are kept. However, there are further differences, which arise from the fact that in ref.~\cite{Duhr:2025zqw} the \MSbar\ renormalisation constants were extracted from the fully off-shell Green's function. The latter is free of infrared (IR) singularities, as the external (off-shell) momentum acts as an infrared regulator~\cite{Poggio:1976qr,PhysRevD.14.2123}. The off-shell two-point function involves logarithms which are singular in the on-shell limit. Below we will discuss how we address this issue. At the same time, while the fully off-shell treatment in ref.~\cite{Duhr:2025zqw} has the advantage that it is free of IR singularities, it comes at the price that one needs to consider additional redundant dimension-six operators that vanish once the equations of motion are imposed. Hence, when working in the on-shell scheme, we do not need to consider additional redundant operators. From this perspective the renormalisation in the on-shell scheme is easier to implement than in the \MSbar-scheme.

Let us now detail how we have performed the computation of the form factors in eq.~\eqref{eq:Sigma1_2} needed to extract the renormalisation constants in the on-shell scheme. We work in dimensional regularisation in $D=4-2\varepsilon$ dimensions~\cite{tHooft:1972tcz}. 
The relevant two-loop diagrams have been generated with \texttt{FeynArts} \cite{hep-ph/0012260}, and the Dirac and colour algebra has been carried out with the use of \texttt{FeynCalc} \cite{2001.04407} and \texttt{FORM} \cite{ 1203.6543,1707.06453}. The Feynman integrals have been organised in integral families and reduced to master integrals (MI) via integration-by-part (IBP) reduction~\cite{Tkachov:1981wb,Chetyrkin:1981qh} with \texttt{Kira} \cite{ 1705.05610,2008.06494}. We note that up to this stage, the computation is identical to the one in the \MSbar-scheme in ref.~\cite{Duhr:2025zqw}, and we refer to that reference for a more detailed discussion. 

\begin{figure}[!h]
    \centering
    \includegraphics[width=0.65\linewidth]{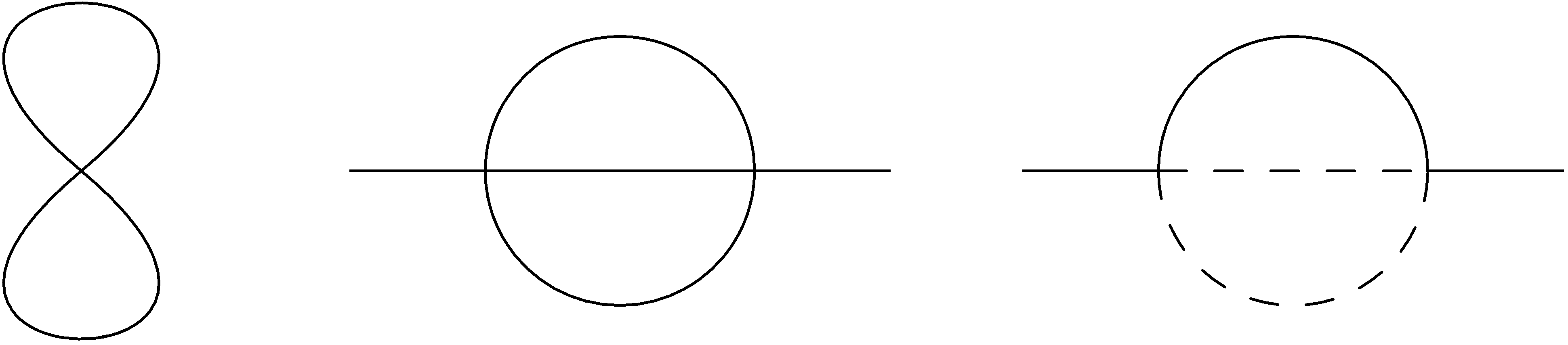}
    \caption{On-shell master integrals for the top quark self-energy at two loops. Solid and dashed lines represent massive and massless propagators, respectively. External lines are on the mass shell, $p^2=M^2$.}
    \label{fig: top-self-MI}
\end{figure}

We now need to impose the renormalisation conditions in eq.~\eqref{eq: OS renormalisation conditions}, and from this point on the computations in the two schemes start to differ. As already mentioned, one important difference with respect to the computation performed in ref. \cite{Duhr:2025zqw} is that we now work in an on-shell environment, with the on-shell condition $p^2 = M^2$ imposed at the integrand level before expansion in the dimensional regulator $\varepsilon$ (because the on-shell renormalisation conditions hold in arbitrary dimensions). In the language of expansions by regions (see, e.g., refs.~\cite{Beneke:1997zp,Smirnov:2002pj,Jantzen:2011nz}), this corresponds to only retaining the \emph{hard region}, where the limit is taken under the integral sign. This changes the reduction to MIs via IBP relations. 
In particular, we obtain a reduced number of MIs compared to the off-shell computation in ref. \cite{Duhr:2025zqw} in the \MSbar\ scheme. For example, in the one-loop case there is a single MI, the one-loop tadpole integral. In the two-loop case we find three MIs, pictorially shown in figure \ref{fig: top-self-MI}, whose expressions on-shell are known in the literature~\cite{hep-ph/9803493, hep-ph/0301170, hep-ph/0202123, Fleischer:1999hp}. Inserting the results for the MIs into the expressions for the form factors, we can immediately extract the on-shell renormalisation constants from the renomalisation conditions in eq.~\eqref{eq: OS renormalisation conditions}. 

We expand the form factors and the renormalisation constants in terms of bare parameters, i.e., we keep both strong coupling constant, as well as the Wilson coefficients, as bare ones. We can write the following expansion for the renormalisation constants,

\begin{equation}
Z_i = 1 + \sum_{L\geq 1} C(\varepsilon, M)^L
\qty(\frac{\alpha_s^0}{4 \pi})^{L-1}
\qty[ \qty(\frac{\alpha_s^0}{4 \pi}) \delta Z_{i,\textrm{QCD}}^{(L)}
+ \frac{1}{16 \pi^2\Lambda^2} \left( \delta Z_{i,G}^{(L)} + \delta Z_{i,tG}^{(L)} \right)
+ \ord\left(\frac{1}{\Lambda^4}\right) ]\,,
\label{eq: renormalisation_factors}
\end{equation}
where $i \in \{t,M,q,3\}$, $\alpha_s^0 = \frac{(g_s^0)^2}{4 \pi}$ is the bare strong coupling constant and we define the normalisation factor
\begin{equation}
    C(\varepsilon,M)=\Gamma(1+\varepsilon) (4 \pi)^\varepsilon M^{-2\varepsilon}\,.
    \label{eq: Ceps constant}
\end{equation}
The QCD renormalisation constants $\delta Z_{i,\textrm{QCD}}^{(L)}$ are known up to four loops~\cite{1502.01030}, and we reproduced the one- and two-loop results as a side product of our computation. In the remainder of this section we only focus on the dimensions-six SMEFT contributions. It is easy to see that the triple-gluon operator $\cO_G$ does not contribute at one loop, and only the chromomagnetic operator $\cO_{tG}$ gives a non-trivial result. We find
\begin{equation}
\begin{split}
\delta Z_{t,G}^{(1)} &\,= \delta Z_{M,G}^{(1)} = 0\,,\\
   \delta Z_{M,tG}^{(1)} &\,=  \frac{g_s^0\,c_{tG}^0\,M \, v}{\sqrt{2}} \CF \frac{4 (2 \varepsilon-3)}{\varepsilon(\varepsilon-1)}\,, \\ \delta Z_{t, tG}^{(1)} &\,=   \frac{g_s^0\,c_{tG}^0\,M \, v}{\sqrt{2}} \CF \frac{2 (2 \varepsilon-3)}{\varepsilon(\varepsilon-1)}\,,
\end{split}
\end{equation}
where $c_{tG}^0$ is the bare Wilson coefficient for $\cO_{tG}$.
 
The extraction of the renormalisation constants at two-loop order is more involved than the one loop case. This is due to the expansion of eq.~\eqref{eq: OS renormalisation conditions}, which also involves derivatives with respect to the mass of the one-loop amplitude. The final results are also more complicated, and can be shown as an expansion in $\varepsilon$. Specifically, we can write
\begin{equation}\begin{split}
    \delta Z_{i, G}^{(2)} &\,=  g_s^0\,c_{G}^0\,M^2\,\CF\, \sum_{n=-2}^\infty \varepsilon^n  \big[ \delta Z_{i, G}^{(2)} \big]_n\,,\\
    \delta Z_{i, tG}^{(2)} &\,=  \frac{g_s^0\,c_{tG}^0\,M \, v}{\sqrt{2}} \,\CF\,\sum_{n=-2}^\infty\varepsilon^n  \big[ \delta Z_{i, tG}^{(2)} \big]_n\,,
    \end{split}
\end{equation}
where $c_{G}^0$ is the bare Wilson coefficient for $\cO_{G}$.
We have computed the coefficients of the expansion in $\varepsilon$ up to the linear term. Here and in the following $n_f=6$ is the total number of quark flavours, including the top quark, so that the number of massless flavours is $n_f-1=5$. For the mass renormalisation counterterms, we find
\begin{align}
\big[\delta Z_{M, G}^{(2)}\big]_{-2} = &\, - 9\, \CA\,,\notag \\  
\big[\delta Z_{M, G}^{(2)}\big]_{-1}= &- \frac{31}{4}\, \CA\,,\notag \\  
\big[\delta Z_{M, G}^{(2)}\big]_{0} = & \, \frac{1}{8} (8 \pi^2-61) \, \CA\,, \notag\\
\big[\delta Z_{M, G}^{(2)}\big]_{1} = & \, \frac{1}{48} \Big(-1065+136 \pi^2+576 \zeta(3)\Big) \, \CA\,,\\
\big[\delta Z_{M, tG}^{(2)}\big]_{-2} = & \, - 84 C_F +34 \, \CA - 4n_f-12 \,,\notag \\  
\big[\delta Z_{M, tG}^{(2)}\big]_{-1} = &\, -148\,C_F +81\, \CA - 6 n_f-70\,, \notag\\   
\big[\delta Z_{M, tG}^{(2)}\big]_{0} = &\, -\big(200+8 \pi^2\big)\,C_F +\Big(112+\frac{8}{3}\pi^2\Big)\, \CA   - 257+24 \pi^2-\Big(1+\frac{8}{3}\pi^2\Big)n_f \,,\notag\\
\big[\delta Z_{M, tG}^{(2)}\big]_{1}= &\,\Big(98-\frac{140}{3} \pi^2-96 \zeta(3)\Big)\,C_F +\Big(-\frac{55}{2}+26 \pi^2+32 \zeta(3)\Big)\, \CA \notag\\ &  -\frac{1667}{2}+\frac{268}{3} \pi^2+480 \zeta(3)-128 \pi^2 \log(2)+  \qty(\frac{77}{2}-4 \pi^2-32 \zeta(3))n_f  \,. \notag
\end{align}
Similarly, the results for the field renormalisation factor of the top quark at two loops are
\begin{align}
\big[\delta Z_{t, G}^{(2)}\big]_{-2} = &\, - \frac{9}{2} \, \CA,\notag \\  
\big[\delta Z_{t, G}^{(2)}\big]_{-1}= &- \frac{3}{2}\, \CA,\notag \\  
\big[\delta Z_{t, G}^{(2)}\big]_{0} = & \, \frac{1}{4} \big(16 \pi^2-31\big) \, \CA, \notag\\
\big[\delta Z_{t, G}^{(2)}\big]_{1} = & \, \frac{1}{8} \big(-535+56 \pi^2+384 \zeta(3)\big) \, \CA\,,\\
\big[\delta Z_{t, tG}^{(2)}\big]_{-2} = & \, - 42 C_F +\frac{77}{4} \, \CA - 2n_f-24 \,,\notag \\  
\big[\delta Z_{t, tG}^{(2)}\big]_{-1} = &\, -95C_F +\frac{449}{8}\, \CA - 7 n_f-38 \,, \notag\\   
\big[\delta Z_{t, tG}^{(2)}\big]_{0} = &\, \Big(\frac{59}{2}-2 \pi^2\Big)C_F +\Big(-\frac{81}{16}+\frac{31}{6}\pi^2\Big)\, \CA - \frac{1}{3}(1245-116 \pi^2)-\frac{1}{6}\big(39+8 \pi^2\big)n_f \,,\notag\\
\big[\delta Z_{t, tG}^{(2)}\big]_{1}= &\,\qty(\frac{4217}{4}-\frac{395}{3} \pi^2-24 \zeta(3))\,C_F +\qty(-\frac{17931}{32}+\frac{827}{12} \pi^2+62 \zeta(3))\, \CA \notag\\ &  -\frac{1}{6}\qty(5715-860 \pi^2-4800 \zeta(3)+1344 \pi^2 \log(2))  \notag\\ &  + \frac{1}{12}\qty(219-56 \pi^2-192 \zeta(3))n_f  \,. \notag
\end{align}
We notice that all the expressions take a gauge-invariant form, as they do not depend on $\xi$. While this is expected for the mass renormalisation factor, in this case it also holds for the field renormalisation constant. The gauge-invariance property of the fermion field renormalisation also holds in pure QCD up to two loops, with the $\xi$-dependence arising at three loops \cite{hep-ph/0005131}.

\subsection{The two-loop relation between the pole and the \texorpdfstring{$\overline{\textrm{MS}}$}{MS} mass}
In the previous subsection we have computed the field and mass renormalisation constants at two loops for the top quark in the on-shell scheme. The same quantities were computed in the \MSbar\ scheme in ref.~\cite{Duhr:2025zqw}.
We can combine these two results to
obtain the relation up to two loops between the pole mass, renormalised in the on-shell scheme, and the one renormalised in the $\overline{\textrm{MS}}$ scheme, that has the property of running with the $\overline{\textrm{MS}}$ renormalisation scale $\mu$. 

The relation between the bare mass $m^0$ and the pole mass $M$ was given in eq.~\eqref{eq:Z_M_def}, while the \MSbar-renormalised mass $m=m(\mu)$ is defined by (cf.~eq.~\eqref{eq:Z_m_matrix_def}),
\beq
m^ 0=Z_m\,m\,.
\eeq
We therefore obtain the relation
\begin{equation}
    M = \frac{Z_m}{Z_M}m(\mu).
\end{equation}
In ref.~\cite{Duhr:2025zqw} the \MSbar\ counterterm was expressed as an expansion in the \MSbar-renormalised coupling, while, as already outlined in the previous subsection, $Z_M$ is expressed in terms of bare parameters, in particular the bare strong coupling constant and the bare Wilson coefficients. In order to consistently derive a  relation, we translate these bare parameters into renormalised ones using the respective $\overline{\textrm{MS}}$ renormalisation factors, noting that only the one-loop renormalisation \cite{1503.08841,1607.05330,Duhr:2025zqw} is required in our case. Additionally, both $Z_M$ and $Z_m$ depend on the renormalised mass $M$ and $m(\mu)$, respectively. We again express everything in terms of the $\overline{\textrm{MS}}$ mass, so that we can write
\begin{equation}
    Z_M = Z_M(M) = Z_M\qty(\frac{Z_m}{Z_M} m(\mu) )\,.
\end{equation}
Expanding up to two loops, we obtain
\begin{eqnarray}
\label{eq:ZmZMexp}
\frac{Z_m}{Z_M} & = & \, \qty(\frac{Z_m}{Z_M})_{\text{QCD}} + \frac{1}{16\pi^2\Lambda^2} \Big(\xi_{G} + \xi_{tG}\Big) + \ord\left(\frac{1}{\Lambda^4}\right)\,.
\end{eqnarray} 
The first term of eq.~\eqref{eq:ZmZMexp} is the relation between the on-shell and $\overline{\textrm{MS}}$ masses in pure QCD, which is known up to four loops \cite{hep-ph/9911434,1502.01030}. In our conventions, we reproduce the corresponding two-loop QCD result. The remaining terms capture the contributions from the operators $\cO_G$ and $\cO_{tG}$. They admit the perturbative expansion
\beq\bsp
\xi_G = {g_s\,c_G}\,m^2\,C_F\,\sum_{L\ge 1}\left(\frac{\alpha_s}{4\pi}\right)^{L-1}\,\xi_G^{(L)}+\ord(\varepsilon)\,,\\
\xi_{tG} = \frac{g_s\,c_{tG}\,m\,v}{\sqrt{2}}\,C_F\,\sum_{L\ge 1}\left(\frac{\alpha_s}{4\pi}\right)^{L-1}\,\xi_{tG}^{(L)}+\ord(\varepsilon)\,,
\esp\eeq
where we dropped the explicit $\mu$ dependence from the $\overline{\textrm{MS}}$-renormalised parameters. We find
\beq\bsp
\xi_G^{(1)} &\,= 0\,,\\
\xi_{G}^{(2)}&\,= \frac{13-8 \pi^2}{8}\,\CA\,, \\   
\xi_{tG}^{(1)}&\,=-4\,,\\
\xi_{tG}^{(2)}&\,= \frac{1}{3}\Big[24(\pi^2+6)\CF -4(67+2 \pi^2)\CA - (5- 8 \pi^2)n_f + 579 - 72 \pi^2\Big]\,.
\label{eq: relation OSMS}
\esp\eeq
 The resulting expression in eq.~\eqref{eq: relation OSMS} is finite, thus providing an important check for the computation, if compared with the $\overline{\textrm{MS}}$ results of ref.~\cite{Duhr:2025zqw}.

\subsection{On-shell renormalisation of the massless quark and gluon fields}
\label{sec: gluon on-shell}

So far we have only discussed the on-shell renormalisation of the top quark, which is the only massive particle in our theory. In this subsection we compute the on-shell renormalisation constants for the gluon field and the massless quarks. We emphasise that, strictly speaking, the on-shell scheme is only defined for massive particles. Nevertheless, it is often useful to extend the on-shell scheme to massless particles. The renormalisation condition is chosen such that the self-energy has a single pole with unit residue at $p^2=0$. The resulting renormalisation constants naturally arise in the computation of scattering amplitudes with massless particles from the LSZ reduction formula. 

Let us start by discussing the massless quarks. We can immediately use all the notations and conventions for the quark self-energy defined in section~\ref{eq:top_OS}, where $m^0$ and $M$ still denote the bare and on-shell masses of the top quark. We define the renormalisation constant $Z_q$ by
\beq
q^0 = \sqrt{Z_q}\,q\,,
\eeq
where $q^0$ and $q$ denote the bare and renormalised massless quark fields, respectively. $Z_q$ admits a perturbative expansion as in eq.~\eqref{eq: renormalisation_factors}, with $i=q$. The QCD contributions from $\delta Z_{q,\textrm{QCD}}$ up to two loops can be found in refs.~\cite{0705.1975, 0707.4139}. We can extract the coefficients at different loop orders in exactly the same fashion as for the massive quark. In this case, the computation is even simpler. Indeed, in dimensional regularisation scaleless loop integrals vanish. It is then easy to see that if we impose the on-shell condition $p^2=0$, only diagrams that contain a closed heavy-quark loop contribute. At one loop all diagrams are scaleless in the on-shell limit. At two loops the triple-gluon operator leads to scaleless integrals, and only the chromomagnetic operator contributes. We find
\beq\bsp
\delta Z_{q,G}^{(1)} &\,= \delta Z_{q,G}^{(2)} = \delta Z_{q,tG}^{(1)} =0 \,,\\
     \delta Z_{q, tG}^{(2)} &\,= \CF \frac{4 M v\,g_s^0\,c_{tG}^0}{\sqrt{2}} \frac{(2 \varepsilon-3)}{\varepsilon(\varepsilon-1)(\varepsilon-2)(2 \varepsilon+1)}\,.
     \esp
\end{equation}

Let us now discuss the on-shell renormalisation of the gluon field. We introduce the renormalisation constant $Z_3$ relating the bare and renormalised gluon fields,
\beq
G_{\mu}^{0,a} = \sqrt{Z_3}\,G_{\mu}^a\,.
\label{eq: gluon renormalisation constant OS}
\eeq
The renormalisation conditions are similar to the quark case. More precisely, 
as a consequence of Lorentz and gauge invariance, in terms of bare quantities, the gluon self-energy takes the form 
\begin{equation}
    \Pi^{\mu\nu}(p,m^0) = (p^2 g^{\mu\nu}-p^\mu p^\nu) \Pi(p, m^0)\,,
\end{equation}
where $m^0$ is the bare top mass.
The on-shell gluon renormalisation factor $Z_3$ is then obtained by requiring that the renormalised self-energy has a pole with unit residue for $p^2=0$. This results in the condition
\begin{equation}
    \frac{Z_3^{-1}}{1-\eval{\Pi(p, Z_M M)}_{p^2=0}} =1\,,
    \label{eq: gluon on shell condition}
\end{equation}
where we used eq.~\eqref{eq:Z_M_def}.

For the extraction of the form factor $\Pi (p,Z_M M)$ we note that the projector employed in ref.~\cite{Duhr:2025zqw} becomes singular in the $p^2=0$ limit. To avoid this issue we introduce an auxiliary light-like momentum $r^\mu$, so that
\begin{equation}
    \eval{\Pi(p, Z_M M)}_{p^2=0} = -\lim_{p^2\rightarrow0}\frac{1}{(p\cdot r)^2} \Pi^{\mu \nu}(p, Z_M M) r_\mu r_\nu\,,
\end{equation}
where the limit is performed at the integrand level, and we avoid explicit $1/p^2$ terms. For the IBP reduction, we treat $r$ as an additional external momentum and we supply the topologies with auxiliary propagators involving $r$ and loop momenta so that we avoid singularities related to the integral reduction. The final result will clearly be independent of $r$.
In our case, there is only one MI, for both the one-loop and two-loop cases, namely massive top tadpole in the former case, and the product of two massive tadpoles in the latter.

The renormalisation constant $Z_3$ admits a perturbative expansion as in eq.~\eqref{eq: renormalisation_factors}, with $i=3$.
After expanding eq.~\eqref{eq: gluon on shell condition} in perturbation theory, we see that the computation also involves derivatives of the form factor with respect to the top mass. Following the conventions of eq.~\eqref{eq: renormalisation_factors} the one-loop result is
\begin{equation}
 \delta Z_{3, tG}^{(1)} = \frac{8\, M\,  v\,g_s^0\,c_{tG}^0}{\sqrt{2} \varepsilon} \,.
\end{equation}
At two loops, we find
\begin{eqnarray}\nonumber
    \delta Z_{3, tG}^{(2)} &=&  -\frac{32\, M\,  v\,g_s^0\,c_{tG}^0}{3 \sqrt{2} \varepsilon^2}  -\CF\frac{16\, M\,  v\,g_s^0\,c_{tG}^0 }{ \sqrt{2}} \frac{(2 \varepsilon^2+\varepsilon-4)}{\varepsilon^2(\varepsilon-2)}\\
    &+& \CA \frac{4\, M\,  v\,g_s^0\,c_{tG}^0}{ \sqrt{2}} \frac{(8 \varepsilon^4-8 \varepsilon^3-16 \varepsilon^2+10 \varepsilon+7)}{\varepsilon^2(\varepsilon-2)(\varepsilon-1)(2\varepsilon+1)}\,, \\ 
\nonumber    \delta Z_{3, G}^{(2)} &=& \CA M^2\,g_s^0\,c_G^0\, \frac{12 (6 \varepsilon^2-11 \varepsilon-1)}{\varepsilon^2(\varepsilon-2)(2\varepsilon-1)(2\varepsilon+1)}\,.
\end{eqnarray}

\section{Gluon field renormalisation with four-quark operators}
\label{sec:4F}

In the previous section and in ref.~\cite{Duhr:2025zqw}, we have computed the contributions of the triple-gluon and chromomagnetic operators to the mass and field renormalisation constants at two-loop order. In order to complete the program of renormalising the quark and gluon self-energies obtained from the Lagrangian in eq.~\eqref{eq:LQCD6}, we also need to determine the contributions from the four-quark operators in eq.~\eqref{eq: four-quark}. Due to the chiral nature of the electroweak gauge interactions, the four-quark operators involve chiral fermions. It is well known that computations with chiral fermions in dimensional regularisation pose some technical challenges, because the $\gamma^5$ matrix is an intrinsically four-dimensional object and has no simple generalisation to arbitrary dimensions. Computations in the SMEFT involving four-fermion operators are therefore challenging. A detailed discussion of these issues related to the renormalisation constant for quarks will be presented in a dedicated publication~\cite{Duhr:2026btw}. It turns out that for the gluon, some of these technical issues can be circumvented, and we therefore present in this section the contributions of four-quark operators to the renormalisation of the gluon field. We start by reviewing the treatment of four-fermion operators with chiral fermions in dimensional regularisation in section~\ref{sec: fourquark}, and we present the computation of the gluon renormalisation constant in the \MSbar\ and on-shell schemes in sections~\ref{sec: gluon renormalisation} and~\ref{sec:4F_gluon_OS}. Note that in this section we consider all the four-quark operators in eq.~\eqref{eq: four-quark}, keeping in mind that we may need to be careful when working in a scheme where the bottom quark is treated as massless (see the discussion at the end of section~\ref{sec:SMEFT_QCD}).

\subsection{Four-quark operators in the BMHV scheme}
\label{sec: fourquark}
In contrast to the operators $ \mathcal{O}_{G}$ and $ \mathcal{O}_{tG}$, the four-fermion operators in eq.~\eqref{eq: four-quark} distinguish between left- and right-handed components of the fermion fields, even when imposing CP-conservation at the Lagrangian level. A crucial consequence is that when computing amplitudes, we need the chiral projectors 
\begin{equation}
    P_{R/L} = \frac{1 \pm \gamma^5}{2}\,.
\end{equation}
This raises the problem of the treatment of $\gamma^5$ in $D=4-2\varepsilon$ dimensions. This is a well-known issue, and a continuation scheme for $\gamma^5$ needs to be adopted. One possible choice for this task is the Naive Dimensional Regularisation (NDR) scheme, which maintains the $\gamma^5$ anti-commutation property with the other gamma matrices in $D$ dimensions. This scheme choice is known to be mathematically inconsistent, as it breaks the cyclicity property of the trace, but it has been proven to give unambiguous results at one-loop, when a consistent reading point for the amplitude is used \cite{Kreimer:1989ke, Korner:1991sx, Kreimer:1993bh, hep-th/0005255, 2012.08506}.

Another continuation scheme for the $\gamma^5$ matrix treatment is the Breitenlohner-Maison-'t Hooft-Veltman (BMHV) scheme \cite{tHooft:1972tcz, Breitenlohner:1977hr}. In this scheme, $\gamma^5$ is treated as a purely four-dimensional object, while the $D$-dimensional space-time is split into a four- and $\varepsilon$-dimensional part, so that we can write 
\begin{equation}
    \gamma_\mu = \Bar{\gamma}_\mu + \hat{\gamma}_\mu, \ \ g_{\mu \nu}=\Bar{g}_{\mu \nu}+ \hat{g}_{\mu \nu}\,,
\end{equation}
where we use the symbols $\Bar{}$ and $\hat{}$ to indicate the four- and $\varepsilon$-dimensional components, respectively. In this context, $\gamma^5$ satisfies
\begin{equation}
    \{\Bar{\gamma}_\mu,\gamma_5\}=0, \ \ [\hat{\gamma}_\mu,\gamma_5]=0\,.
    \label{eq: ga5relations}
\end{equation}
The resulting algebra has been proven to be mathematically consistent at all loop orders \cite{Breitenlohner:1977hr, Speer:1974cz, Breitenlohner:1975hg, Breitenlohner:1976te, Costa:1977pd, Aoyama:1980yw}. Since we want to perform a computation involving traces of $\gamma$ matrices at two-loop order, we make the choice of the BMHV scheme, which ensures mathematical consistency.

In the BMHV scheme, when we construct a $D$-dimensional four-quark physical operator, it will automatically be projected into the four-dimensional space due to the identity
\begin{equation}
    P_R \gamma^\mu P_L = \Bar{\gamma}^\mu P_L\,.
\end{equation}
However, as an infinite-dimensional representation of the Dirac algebra is required also for the four-dimensional subspace in the BMHV scheme, the Fierz identities do not hold \cite{Collins:1984xc, 2310.13051}, and a minimal basis of operators obtained by employing such identities is defined up to a set of evanescent operators \cite{2211.09144}. The basis of third-generation operators defined in eq.~\eqref{eq: four-quark} is one possible choice of minimal basis, and when relating this basis with the conventions of, e.g., ref.~\cite{1008.4884} one should take into account evanescent structures appearing when Fierz transformations are applied to the operators.\footnote{For a discussion of the modifications of the multiloop renormalisation group equations made necessary by the presence of evanescent operators, see ref.~\cite{Schubert:1988ke}.}

The evanescent operators of the Fierz type are not the only ones that appear in the BMHV scheme, as the algebra defined by the continuation of $ \gamma^5 $ introduces a potentially infinite set of new evanescent structures involving the $\hat{\gamma}$  matrices~\cite{2310.13051}. However, the number of such structures generated at one loop is limited and depends on the scheme adopted for continuing the fermion Lagrangian to $D$ dimension. In our case, this is determined by the treatment of the gluon-fermion vertex, and specifically, whether the interaction is continued to $D$ dimension or kept strictly four-dimensional. This will be discussed in detail in section~\ref{sec: gluon renormalisation}.

Finally, it is worth mentioning that continuing the fermion kinetic term to $D$ dimension naturally breaks chiral gauge symmetry, which can be restored through the introduction of finite counterterms~\cite{hep-th/9905076, hep-th/0209023, 2004.14398, 2109.11042, 2208.09006, 2303.09120, 2312.11291, 2205.10381, 2406.17013, 2507.19589,Kuhler:2025znv,vonManteuffel:2025swv}. In our case, since the electroweak sector is treated as static and we focus on QCD corrections, the breaking affects a global chiral symmetry. We will comment on this below.

\subsection[Two-loop gluon self-energy in the MS scheme]{Two-loop gluon self-energy in the $\overline{\text{MS}}$ scheme}
\label{sec: gluon renormalisation}

\begin{figure}[!h]
    \centering
    \includegraphics[width=0.65\linewidth]{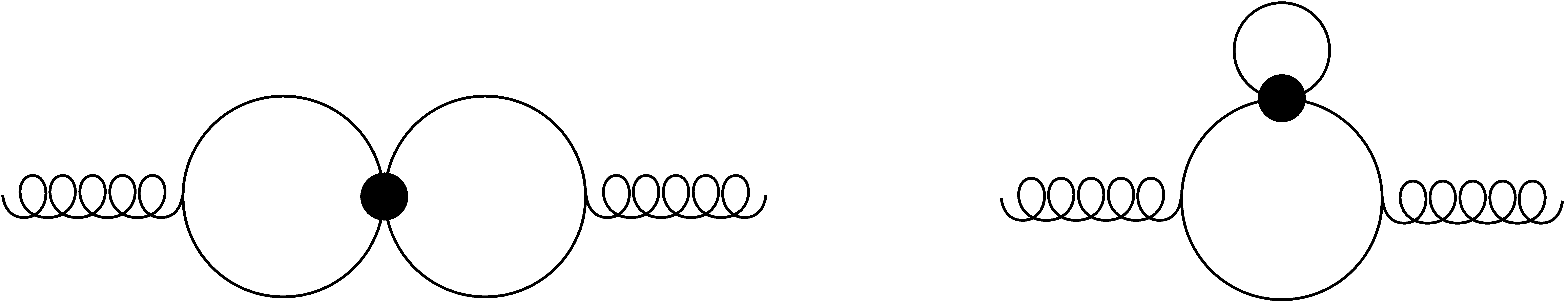}
    \caption{Two-loop diagrams of the gluon self-energy with an insertion of four-quark operators (in black).}
    \label{fig: gluon-self-diagrams}
\end{figure}

We now discuss the gluon renormalisation constant $Z_G$\footnote{This differs from the renormalisation constant in the on-shell scheme, which we denoted with $Z_3$, cf. equation \eqref{eq: gluon renormalisation constant OS}.} in the \MSbar\ scheme. It admits the SMEFT expansion,
\begin{equation}
\label{eq:Z_G_def}
     Z_G = 1 + \sum_{L\geq 1}\qty(\frac{\alpha_s}{4 \pi})^{L-1}\Bigg[\qty(\frac{\alpha_s}{4 \pi}) \delta Z_{G,\textrm{QCD}}^{(L)} + \frac{1}{16 \pi^2\Lambda^2}\,\left(\delta Z_{G,G}^{(L)} + \delta Z_{G,tG}^{(L)} + \delta Z_{G,4q}^{(L)}\right)\, +\ord\left(\frac{1}{\Lambda^4}\right) \Bigg]\,,
\end{equation}
where $\delta Z_{G,\textrm{QCD}}$ are the pure-QCD contributions, known up to five  loops~\cite{1709.08541}, and $\delta Z_{G,G}$ and $\delta Z_{G,tG}$ capture the effects of the triple-gluon operator $\cO_{G}$ and the chromomagenatic operator $\cO_{tG}$, whose two-loop expressions were obtained in ref.~\cite{Duhr:2025zqw}. The term $\delta Z_{G,4q}$ arises from four-quark operators through two-loop Feynman diagrams like those shown in figure~\ref{fig: gluon-self-diagrams}, and it is one of the main results of this section.

As we see from figure \ref{fig: gluon-self-diagrams}, the subgraphs related to the two-loop amplitude are the one-loop quark self-energy and the one-loop corrections to the quark gluon vertex. The renormalisation of the former has already been discussed in section~\ref{sec:SMEFT_QCD}, where the one-loop renormalisation factors of the quark masses was shown, and used as an argument to consider both the top and bottom quark as massive.

An important consequence of the diagrams in figure~\ref{fig: gluon-self-diagrams} is that evanescent operators containing four quark fields do not contribute to the computation, as the two-loop gluon self-energy is not sensitive to them. In a renormalised framework, such evanescent operators are generated at one loop when computing corrections to the four-quark vertex. However, in the case of the gluon self-energy, four-fermion operators enter for the first time at the two-loop level, and therefore there are no subdiagrams involving one-loop corrections to the four-quark vertex. As a result, the amplitude is independent of insertions from evanescent operators involving four quark fields, such as the Fierz-type structures. This argument does not apply to evanescent operators involving two fermions and a gluon field, which may contribute and can be explicitly generated through the one-loop renormalisation of the gluon-fermion vertex. This is discussed in more detail in appendix \ref{sec: penguin appendix}.

As discussed in the literature \cite{Buras:1989xd, Dugan:1990df, hep-ph/9412375,hep-ph/9806471, 2211.09144}, insertions of these operators in one-loop counterterm diagrams need to be taken into account. However, in our specific case of the gluon self-energy, we checked that the insertions of such evanescent operators (listed in table~\ref{tab: penguin CT evanescent} of appendix \ref{sec: penguin appendix}) in relevant one-loop counterterm diagrams contribute only to the evanescent sector.\footnote{We remark that this is specific to the computation we performed, and, for example, does not apply for the quark self-energy where these operators contribute to the physical sector via one-loop diagrams, cf., e.g., ref.~\cite{Duhr:2026btw}.} Such operators would only be relevant for the renormalisation of the self-energy starting at three loops, while their contributions vanish in the $D\rightarrow4$ limit at the two-loop level. Therefore, in our case, this implies that only the counterterms of divergent physical and redundant structures are needed for the renormalisation procedure.\footnote{We do not generate any chiral-symmetry breaking four-dimensional structures, because the computation is performed in the broken phase, where chiral symmetry is broken by the quark mass term.} These are summarised in table~\ref{tab: penguin CT} of appendix \ref{sec: penguin appendix}.

\begin{figure}[!h]
    \centering
    \includegraphics[width=0.65\linewidth]{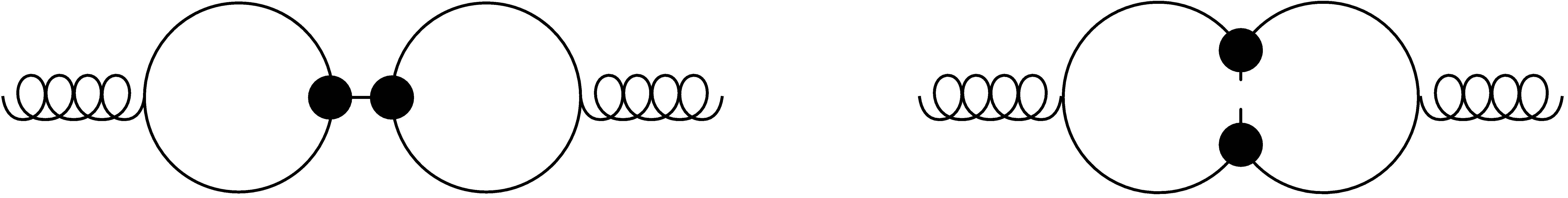}
    \caption{Possible connections for one of the factorised diagram topologies when four-quark structures are inserted in the gluon self-energy. On the left, we show the reducible connection, while on the right we have the 1PI diagram.}
    \label{fig: gluon-self-factorised}
\end{figure}

Once the one-loop off-shell renormalisation for relevant subgraphs is completed, we can focus on the computation of the two-loop diagrams. As we previously mentioned, the presence of factorised diagrams simplifies the computation. The topologies needed to organise the two-loop integrals are products of one-loop topologies and, as a result, the amplitude will be only expressed in terms of products of one-loop integrals, in this case tadpole and bubble type. 
Although the factorised form of the two-loop diagrams in figure~\ref{fig: gluon-self-diagrams} may suggest that they might not be One Particle Irreducible (1PI), and can be obtained by stitching together one-loop diagrams, this is not the case, due to the way fermion lines are connected. In particular, one possible connection renders the diagram effectively reducible, expressed as a product of two traces and giving no contribution to the final pole. However, the other connection leads to one single Dirac trace and is 1PI. This can be visualised by unlocking the four-quark vertex via an auxiliary field as shown in figure \ref{fig: gluon-self-factorised}. The 1PI case corresponds to the auxiliary propagator connecting two internal fermion propagators. 

Following the R-operation~\cite{Zimmermann:1969jj,Bogoliubov:1957gp,Hepp:1966eg}, we consider the one-loop diagrams obtained by inserting relevant counterterm operators. The remaining pole terms have a very simple structure and can be cast in the form 
\begin{align}
\frac{2 g_s^2}{9 (16\pi^2)^2 \Lambda^2 \varepsilon^2} 
&\Big(
  2(c_{Qt}^{(8)} + c_{Qb}^{(8)} + c_{tt} + c_{bb}) 
  + c_{tb}^{(8)} 
  + 2c_{QQ}^{(1)} - (\CA - 2\CF - 2)c_{QQ}^{(8)} 
\Big) T_6^{\mu\nu} \notag \\
&\quad + \frac{2 g_s^2}{(16\pi^2)^2 \Lambda^2 \varepsilon^2} 
m_t m_b 
\left(
  2 c_{QtQb}^{(1)} + (2 \CF - \CA)c_{QtQb}^{(8)}
\right) T_4^{\mu\nu}\,,
\label{eq: pole gluon}
\end{align}
where $m_t$ and $m_b$ are the top and bottom masses in the \MSbar\ scheme and $T_4^{\mu \nu}$ and $T_6^{\mu \nu}$ are the tensor structures defined by
\begin{equation}
    T_4^{\mu \nu} = p^2 g^{\mu\nu}- p^\mu p^\nu \textrm{~~~~and~~~~} T_6^{\mu \nu} = p^2 (p^2 g^{\mu\nu} - p^\mu p^\nu)\,.
\end{equation}
The first line of eq.~\eqref{eq: pole gluon}, proportional to the tensor structure $T_6^{\mu \nu}$, is renormalised by the dimension-six class II (i.e., vanishing after the on-shell equations of motion are imposed) operator
\begin{equation} 
    \cO_{\partial^2 G} = (D^\mu G^a_{\mu \nu} - g_s \bar \Psi_i \bar \gamma_\nu T^a \Psi_i)^2\,,
\end{equation}
resulting in the two-loop counterterm
\begin{equation}
    \delta Z_{\partial^2G}^{(2)}=-\frac{ 1}{9 \varepsilon^2}  
    \Big(2(c_{Qt}^{(8)}+c_{Qb}^{(8)}+c_{tt}+c_{bb})+c_{tb}^{(8)}+2c_{QQ}^{(1)}-(\CA-2\CF-2)c_{QQ}^{(8)}\Big)\,,
    \label{eq:Zd2G}
\end{equation}
where we matched the conventions introduced in eq.~\eqref{eq:Z_G_def}. 
The second line of eq.~\eqref{eq: pole gluon} contributes to the gluon renormalisation constant $Z_G$ at two loops, and we have
\begin{equation}
    \delta Z_{G,4q}^{(2)} = \frac{2  \, m_t \, m_b}{\varepsilon^2}\bigg(2 c_{QtQb}^{(1)} - (\CA-2 \CF)c_{QtQb}^{(8)}\bigg)\,.
    \label{eq:ZG4q}
\end{equation}
This is our final result for the four-quark contribution to the gluon field renormalisation constant in the \MSbar\ scheme. Note that the expressions in eqs.~\eqref{eq:Zd2G} and~\eqref{eq:ZG4q} only have a double pole in $\varepsilon$, but the single pole is absent. This has an important consequence, which we discuss in the remainder of this section.

It is well known that in QCD the renormalisation and the running of the strong coupling are tightly linked to the renormalisation of the gluon self-energy. We can thus use our computation to obtain the contribution of four-quark operators to the two-loop QCD beta function in the SMEFT. 
We observe that the computations in the covariant and background gauges are identical, as no gluon vertices are involved. Therefore, it is straightforward to also derive the contribution of four-quark operators to the counterterm for the strong coupling by the relation $Z_{g_s} \sqrt{Z_G} =1$. We get
\begin{equation}
    \delta Z_{g_s,4q}^{(2)} =-\frac{ m_t \, m_b}{ \varepsilon^2}\Big(2 c_{QtQb}^{(1)} - (\CA-2 \CF)c_{QtQb}^{(8)}\Big)\,,
    \label{eq: strong coupling counterterm}
\end{equation}
where we match again the conventions introduced in eq.~\eqref{eq:Z_G_def}. 

The lack of a single pole implies that none of the four-quark operators of eq.~\eqref{eq: four-quark} contribute to the two-loop running of $\alpha_s$.\footnote{More generally, from this computation we can infer that none of the four-fermion operators contribute to the two-loop running of $\alpha_s$, as the gluon self-energy is not sensitive to the flavour structure of the inserted four-fermion operators, but only to the Dirac and colour structure, and the operators in eq.~\eqref{eq: four-quark} comprise all the structures of a SMEFT basis.} This is in agreement with, and verifies independently, the arguments presented in ref.~\cite{2308.06315} for factorisable diagrams, such as the ones in figure~\ref{fig: gluon-self-diagrams}. In such contributions, the single pole vanishes up to finite shifting, with non-zero contributions arising for the double pole and finite terms. This result also matches the recent findings of ref.~\cite{2507.10295}, where SMEFT four-top operators were considered, and the LEFT computation of ref.~\cite{2507.08926}. The absence of the single pole is a non-trivial check for our result for the gluon renormalisation constant. As another strong check of our computation, we 
mention that the counterterm in eq.~\eqref{eq: strong coupling counterterm} is expected by the finiteness of the two-loop running of $\alpha_s$. Specifically, this is due to the mixing of the operators $\cO_{QtQb}^{(R)}$, $R\in\{1,8\}$ into the chromomagnetic dipole operators, which contribute to the running of the strong coupling already at one loop \cite{1503.08841, 1607.05330}. Equation~\eqref{eq: strong coupling counterterm} provides the necessary term to cancel the effect of this mixing in the running of $\alpha_s$ at two loops. 

We conclude this discussion by stressing that, together with the results of ref.~\cite{Duhr:2025zqw}, our computation completes the running of the strong coupling at two loops within the QCD sector of the SMEFT if only the top quark is considered massive. This running should be used in all computations in the SMEFT with dimension-six operators. We note however, that one often uses automated tools, like matrix element generators combined with parton shower codes, which have their own implementation of the running of the strong coupling, and this running is often restricted to the SM. To remedy such a potential mismatch in the computation we define a decoupling scheme in appendix \ref{sec:decoupling}. In analogy with the decoupling of the heavy quarks from the running of $\alpha_s$ in the SM~\cite{Bernreuther:1981sg},  we define an effective coupling constant that has the same running as the strong coupling constant in the SM, so that the SMEFT has effectively decoupled from the running of the strong coupling.

\subsection{Two-loop gluon self-energy in the on-shell scheme}
\label{sec:4F_gluon_OS}
Following the computation of section \ref{sec: gluon on-shell}, we can extend our results to the renormalisation of the gluon field in the on-shell scheme by also including the four-fermion operators of eq.~\eqref{eq: four-quark}. As in this computation both top and bottom quarks are massive, we first renormalise the masses in the on-shell scheme at one-loop. Letting $m_t^0 = Z_{M_t} M_t$ and $m_b^0 = Z_{M_b} M_b$, the four-quark operators lead to the corrections
\begin{equation}
\begin{split}
    &  \delta Z_{M_t,4q}^{(1)} = C(\varepsilon, M_t) \,4 M_t^2 \frac{c_{Qt}^{(1)}+\CF c_{Qt}^{(8)}}{ \varepsilon(\varepsilon-1)} - C(\varepsilon, M_b) \, \frac{M_b^3}{M_t} \frac{(2 \CA+1)c_{QtQb}^{(1)}+\CF c_{QtQb}^{(8)}}{  \varepsilon (\varepsilon-1)}\,, \\ &\delta Z_{M_b,4q}^{(1)} = C(\varepsilon, M_b) \,4 M_b^2 \frac{c_{Qb}^{(1)}+\CF c_{Qb}^{(8)}}{\varepsilon(\varepsilon-1)} - C(\varepsilon, M_t) \, \frac{M_t^3}{M_b} \frac{(2 \CA+1)c_{QtQb}^{(1)}+\CF c_{QtQb}^{(8)}}{  \varepsilon (\varepsilon-1)}\,,
\end{split}
\end{equation}
where $C(\varepsilon, M)$ is defined in eq.~\eqref{eq: Ceps constant}, and we follow the notations and conventions of eq.~\eqref{eq: renormalisation_factors}.  These one-loop results agree with ref.~\cite{1512.02508}, and we notice the same mixing pattern discussed in section~\ref{sec:SMEFT_QCD}. 

We can now apply the condition in eq.~\eqref{eq: gluon on shell condition} to extract the gluon renormalisation factor in the on-shell scheme. The computation follows the methodology of section \ref{sec: gluon on-shell}, where we now have additional master integrals due to the mass of the bottom quark, and specifically products of top and bottom tadpoles. The final result reads
\begin{equation}
\begin{split}
    \delta Z_{3,4q}^{(2)}  = & \, C(\varepsilon,M_t)C(\varepsilon,M_b)\frac{2 \, M_t \, M_b}{\varepsilon^2}\bigg(2 c_{QtQb}^{(1)} - (\CA-2 \CF)c_{QtQb}^{(8)}\bigg)\,,
    \label{eq: gluon on-shell 4F}
\end{split}
\end{equation}
where we again follow the notations and conventions of eq.~\eqref{eq: renormalisation_factors}. We only have a double pole contribution from the operators $\cO_{QtQb}^{(R)}$, $R\in\{1,8\}$; the purely left- and right-handed four-quark operators $\cO_{QQ}^{(R)}$, $\cO_{tt}$ and $\cO_{bb}$ do not contribute to $Z_3$ at this order. We note that, since the two-loop contribution is the first contribution of the four-quark operators to the gluon renormalisation factors, expressing eq.~\eqref{eq: gluon on-shell 4F} in terms of bare or renormalised parameters for $g_s$ and the Wilson coefficients is equivalent up to terms of three-loop order, which we neglect. 

\section{Conclusions}
\label{sec: conclusions}
In this work we took another step towards the two-loop renormalisation of fields and masses in the QCD sector of the SMEFT. We have computed the renormalisation factors of the top quark field at two loops in the on-shell scheme in the presence of the dimension-six chromomagnetic dipole moment and triple-gluon operators. Our results complement our previous computation in the $\overline{\text{MS}}$ scheme. Our computation employed standard multiloop techniques and allowed us to extract also the relationship between the on-shell and $\overline{\text{MS}}$ top masses. We also computed the renormalisation of the gluon and light quark fields in the on-shell scheme. Our results are essential ingredients for any two-loop computation in the QCD sector of the SMEFT. 

We also extended our operator basis to include four-quark operators. Four-quark operators due to their chiral nature introduce additional complications in the multiloop computation. In particular, the complexity of treating $\gamma_5$ beyond four dimensions has to be addressed. We extract the two-loop renormalisation factor for the gluon field in the presence of four-fermion operators, taking both top and bottom quarks to be massive. We find that there is no single pole contribution. From the renormalisation constant for the gluon we can extract also the renormalisation of the strong coupling constant. The absence of a single pole means that the four-fermion operators do not contribute to the running of the strong coupling constant, a result consistent with observations in the literature. 

\section*{Acknowledgements}
GV thanks Victor Miralles and Hesham El Faham for useful discussions. We also thank Andres Vasquez for collaboration during the early stages of this project. GV and EV are supported by the European Research Council (ERC) under the European
Union’s Horizon 2020 research and innovation programme (Grant agreement No. 949451) and by a Royal Society University Research Fellowship through grant URF/R1/201553. The work of CD is supported by the DFG project 499573813 “EFTools”.

\appendix

\section{Penguin diagram in BMHV}
\label{sec: penguin appendix}
We focus here on the renormalisation of the quark-gluon vertex at one loop with four-fermion operators in the BMHV scheme. The diagrams are the well-known penguin diagrams, which we compute in the non-anticommuting $\gamma^5$ scheme. The generated UV-divergent tensor structures include four-dimensional physical and redundant structures, as well as evanescent ones. 

Focusing on the divergent four-dimensional structures, the results of the renomalisation procedure are in table \ref{tab: penguin CT}. Specifically, three redundant, class II operators are generated, i.e., $\cO_{QQ\partial G}, \cO_{tt\partial G}$ and $\cO_{bb\partial G}$, as well as the two chromomagnetic dipole operators of up and down type $\cO_{tG}$ and $\cO_{bG}$. The latter set is only generated by $\cO_{QtQb}^{(R)}$, $R\in\{1,8\}$, which mixes with the dipole operators, but does not mix with other four-fermion operators \cite{1312.2014}.

\renewcommand{\arraystretch}{2.2}
\begin{table}[h!]
\centering
\resizebox{0.9\textwidth}{!}{\begin{tabular}{|c|c|}
\hline
Operators generated at one-loop & Counterterm\\
\hline \hline
$\cO_{QQ\partial G} = \qbar \bar \gamma^\nu T^a Q (D^\mu G^a_{\mu \nu}-g_s  \bar \Psi_i \bar \gamma_\nu T^a \Psi_i)$ & $\dfrac{1}{3}(c_{Qt}^{(8)}+c_{Qb}^{(8)}+2 c_{QQ}^{(1)}-(\CA-2\CF-2)c_{QQ}^{(8)}) $
\\

$\cO_{tt\partial G} = \tbar \bar \gamma^\nu T^a t (D^\mu G^a_{\mu \nu}-g_s  \bar \Psi_i \bar \gamma_\nu T^a \Psi_i)$ & $\dfrac{1}{3}(2c_{Qt}^{(8)}+c_{tb}^{(8)}+4c_{tt}) $ \\ 
$\cO_{bb\partial G} = \bbar \bar \gamma^\nu T^a b (D^\mu G^a_{\mu \nu}-g_s  \bar \Psi_i \bar \gamma_\nu T^a \Psi_i)$ & $\dfrac{1}{3}(2c_{Qb}^{(8)}+c_{tb}^{(8)}+4c_{bb}) $ \\ [1.3ex] \hline \hline
$\cO_{tG} = i\,\qbar \bar \tau^{\mu \nu} T^a \Tilde{\phi}\, t\, G_{\mu \nu}^a + \text{h.c.}$ & $\dfrac{m_b\sqrt{2}}{4\, v}((\CA-2\CF)c_{QtQb}^{(8)}-2c_{QtQb}^{(1)})$ \\ $\cO_{bG} = i\,\qbar \bar \tau^{\mu \nu} T^a \phi\, b\, G_{\mu \nu}^a + \text{h.c.}$ & $\dfrac{m_t \sqrt{2}}{4 \, v}((\CA-2\CF)c_{QtQb}^{(8)}-2c_{QtQb}^{(1)})$ \\ [1.3ex] \hline
\end{tabular}}
\caption{Operators generated at one loop for the renormalisation of the off-shell penguin diagram. The first set of operators is redundant, vanishing by the equations of motion. The second set is physical and includes the chromomagnetic dipole operators, and $\Psi_i = \{Q, t, b \}$. The sum over the generation index $i$ is understood. A factor of $\dfrac{g_s}{16 \pi^2 \epsilon}$ for the counterterms is also understood.}
\label{tab: penguin CT}
\end{table}
\normalsize

As it is well known \cite{Gilman:1979bc, Georgi:1984zwz,1308.2627}, the set of redundant operators generated by the penguin diagrams play a crucial role in the renormalisation of four-fermion operators, because of the four-quark structure generated via the classical equations of motion, triggering additional mixing that is not computed via the 1PI one-loop corrections to the four-quark vertex.

When computing Green's functions at one loop in the BMHV scheme, in addition to the divergent four-dimensional tensor structures which determine the $\overline{\text{MS}}$ counterterms, we can generate finite mixing into four-dimensional structures that can break chiral gauge symmetry, as well as divergent evanescent structures. Both are of UV origin and involve $\hat \gamma$ matrices and their contractions. 
In our specific case, we do not find any of the four-dimensional chiral breaking structures, but we generate UV-divergent evanescent structures. As we mentioned in section \ref{sec: fourquark}, the number of such evanescent structures depends on the particular scheme employed when continuing the fermion Lagrangian to $D$ dimensions. Specifically, in our case this depends on whether we consider the gluon-quark interaction to be $D$-dimensional or strictly four-dimensional. In the former case, additional $\hat \gamma$ matrices are involved in the computation, leading to a larger set of evanescent operators. The operators generated in our case, in both schemes, are shown in table \ref{tab: penguin CT evanescent}. The counterterms to these operators in the BMHV scheme have been also computed in the context of the LEFT in ref.~\cite{2310.13051}, and we agree with their results for the overlapping operators.

\renewcommand{\arraystretch}{2.2}
\begin{table}[h!]
\centering
\resizebox{ \textwidth}{!}{\begin{tabular}{|c|c|}
\hline
EV. operators generated at one-loop & Counterterm\\ [1.2ex]
\hline \hline
$\mathcal{E}_{t G}^{R} = \bar t_{R} i \hat \gamma^\mu \overline{\gamma}^\nu
 T^a t_{R} G_{\mu \nu}^a$ & $ 4 m_t \, c_{tt} $
\\
$\mathcal{E}_{b G}^{R} = \bar b_{R} i \hat \gamma^\mu \overline{\gamma}^\nu
 T^a b_{R} G_{\mu \nu}^a$ & $ 4 m_b \, c_{bb}$
\\
$\mathcal{E}_{t G}^{L} = \bar t_{L} i \hat \gamma^\mu \overline{\gamma}^\nu
 T^a t_{L} G_{\mu \nu}^a$ & $ m_t(2\, c_{QQ}^{(1)}+ (2 \CF-\CA) c_{QQ}^{(8)})$
\\
$\mathcal{E}_{b G}^{L} = \bar b_{L} i \hat \gamma^\mu \overline{\gamma}^\nu
 T^a b_{L} G_{\mu \nu}^a$ & $  m_b(2\, c_{QQ}^{(1)}+ (2 \CF-\CA) c_{QQ}^{(8)})$
\\

$\mathcal{E}_{t GD}^{LR} = \bar t_L \hat{\gamma}^\nu T^a t_R D^\mu G_{\mu \nu}^a + \text{h.c.} $ & $\dfrac{1}{3}(4 c_{Qt}^{(1)}+ c_{QtQb}^{(1)})-\dfrac{1}{6}(4 c_{Qt}^{(8)}+c_{QtQb}^{(8)})(\CA-2\CF)$ \\
$\mathcal{E}_{b GD}^{LR} = \bar b_L \hat{\gamma}^\nu T^a b_R D^\mu G_{\mu \nu}^a + \text{h.c.} $ & $\dfrac{1}{3}(4 c_{Qb}^{(1)}+ c_{QtQb}^{(1)})-\dfrac{1}{6}(4 c_{Qb}^{(8)}+c_{QtQb}^{(8)})(\CA-2\CF) $ \\ [1.3ex] \hline \hline
$\mathcal{E}_{tG}^{LR} = i \bar t_{L}  \hat \tau^{\mu \nu}  T^a t_{R} G_{\mu \nu}^a + \text{h.c.}$ & $m_t\left(2 c_{Qt}^{(1)}+(2 \CF -\CA)c_{Qt}^{(8)}\right) -\dfrac{m_b}{4}\left(2 c_{QtQb}^{(1)}+(2 \CF -\CA)c_{QtQb}^{(8)}\right) $  \\
$\mathcal{E}_{bG}^{LR} = i \bar b_{L}  \hat \tau^{\mu \nu}  T^a b_{R} G_{\mu \nu}^a + \text{h.c.}$ & $ m_b\left(2 c_{Qb}^{(1)}+(2 \CF -\CA)c_{Qb}^{(8)}\right) -\dfrac{m_t}{4}\left(2 c_{QtQb}^{(1)}+(2 \CF -\CA)c_{QtQb}^{(8)}\right)$  \\[1.3ex] \hline
\end{tabular}}
\caption{Evanescent operators generated at one loop for the renormalisation of the off-shell penguin diagram. The first set of operators is generated in both the four- and $D$-dimensional schemes for the gauge-fermion vertex. The second set is only generated in the latter. $t$ and $b$ are the physical top and bottom fields, and $\hat \tau^{\mu \nu} = \frac{1}{2} \comm{\hat \gamma^\mu}{\hat \gamma^\nu}$. A factor of $\frac{g_s}{16 \pi^2 \epsilon}$ for the counterterms is understood.}
\label{tab: penguin CT evanescent}
\end{table}
\normalsize

\section{A SMEFT decoupling scheme for the strong coupling}
\label{sec:decoupling}

In our previous work \cite{Duhr:2025zqw}, we extracted the contributions of two SMEFT operators to the running of the strong coupling constant in the $\overline{\textrm{MS}}$-scheme, and in section \ref{sec:4F} we have verified that four-fermion operators do not contribute to the running at two loops.  In general the running of the strong coupling in the SMEFT takes the form
\beq\label{eq:a_s_RGE}
\frac{\rd\alpha_s}{\rd\!\log\mu^2} = \beta_4(\alpha_s)\alpha_s + \beta_6(\alpha_s)\alpha_s + \ord\left(\frac{1}{\Lambda^4}\right)\,,
\eeq
where $\beta_4$ is the SM QCD beta function and $\beta_6$ captures the contributions from dimension-six operators in the SMEFT. They admit the perturbative expansion
\beq\bsp
\beta_4(\alpha_s) &\,= -\sum_{L=1}^{\infty}\beta_{L-1}\left(\frac{\alpha_s}{4\pi}\right)^L\,,\\
\beta_6(\alpha_s) &\,= \sum_{k}\frac{C_k}{\Lambda^2}\beta_{6,k}(\alpha_s) = \sum_{k}\sum_{L=1}^{\infty}\beta_{6,k}^{(L-1)}\,\frac{C_k}{\Lambda^2}\,\left(\frac{\alpha_s}{4\pi}\right)^L\,,
\esp\eeq
and $C_k$ denote the renormalised Wilson coefficients, which satisfy the renormalisation group equation (RGE),
\beq
\frac{\rd C_k}{\rd\! \log\mu^2} = \sum_{m}\gamma_{km}(\alpha_s)\,C_m = \sum_{m}\sum_{L=1}^{\infty}\gamma_{km}^{(L-1)}\left(\frac{\alpha_s}{4\pi}\right)^L\,C_m\,.
\eeq

The RGE in eq.~\eqref{eq:a_s_RGE} should be used in all the computations in the SMEFT with dimension-six operators. However, one often uses automated tools, like matrix element generators or parton shower codes, which have their own implementation of the running of the strong coupling, and this running is often restricted to the SM. Similarly, the strong coupling constant is often extracted together with parton distribution functions (PDFs), which are extracted using SM running. This may then introduce a mismatch in the calculation.

It is well known how to remedy this issue in the case of heavy quarks. Consider for example the SM with $n_f=6$ quark flavours, but in the situation where only the top-quark is considered massive, and the remaining quarks are considered massless. It is then possible to define an effective coupling constant  that has the same running as the strong coupling constant in a theory with $n_f-1=5$ massless quark flavours, and the effective quark has effectively been decoupled form the running of the strong coupling.

Here we propose a similar decoupling scheme for the SMEFT. More precisely, we define an effective coupling $\alpha_{\textrm{eff}}$ that agrees with the SM in the limit $\Lambda\to\infty$,
\beq
\alpha_{\textrm{eff}} = \alpha_s + \ord\left(\frac{1}{\Lambda^2}\right)\,,
\eeq
and has the same RGE as the strong coupling in the SM,
\beq
\frac{\rd\alpha_{\textrm{eff}}}{\rd\!\log\mu^2} = \beta_4(\alpha_{\textrm{eff}})\,\alpha_{\textrm{eff}}\,.
\eeq
These two conditions fix the effective coupling uniquely. We write
\beq
\alpha_{s} = \xi(\alpha_{\textrm{eff}})\,\alpha_{\textrm{eff}} = \alpha_{\textrm{eff}}\left[1+\sum_k\frac{C_k}{\Lambda^2}\,\chi_k\big(\alpha_{\textrm{eff}}\big)+\ord\left(\frac{1}{\Lambda^4}\right)\right]\,,
\eeq
with
\beq
\chi_k(\alpha_{\textrm{eff}}) = \sum_{L=0}^{\infty}\chi_k^{(L)}\,\left(\frac{\alpha_{\textrm{eff}}}{4\pi}\right)^L\,.
\eeq
We find
\begin{align}
\nonumber\chi_k^{(0)} &\,=\frac{\beta_{6,k}^{(0)}}{\Gamma_k^{(0)}+\beta_0}\,,\\
\chi_k^{(1)} &\,=\frac{\beta_{6,k}^{(1)}}{\Gamma_k^{(0)}}-\frac{\beta_{6,k}^{(0)}\big(\Gamma_k^{(1)}+2\beta_1\big)}{\Gamma_k^{(0)}\big(\Gamma_k^{(0)}+\beta_0\big)}\,,\\
\nonumber\chi_k^{(2)} &\,=\frac{\beta_{6,k}^{(2)}}{\Gamma_k^{(0)}-\beta_0}-\frac{\beta_{6,k}^{(1)}\big(\Gamma_k^{(1)}+\beta_1\big)}{\Gamma_k^{(0)}\big(\Gamma_k^{(0)}-\beta_0\big)}-
\frac{\beta_{6,k}^{(0)}\big(\Gamma_k^{(0)}\Gamma_k^{(2)}+3\beta_2\Gamma_k^{(0)}-3\beta_1\Gamma_k^{(1)}-\Gamma_k^{(1)2}-2\beta_1^2\big)}{\Gamma_k^{(0)}\big(\Gamma_k^{(0)}-\beta_0\big)\big(\Gamma_k^{(0)}+\beta_0\big)}\,,
\end{align}
where, assuming diagonal mixing, $\gamma_{km}=\delta_{km}\Gamma_k$ (otherwise the $\chi_k$ solve the corresponding linear system), i.e.
\beq
\Gamma_k^{(L)} = \gamma^{(L)}_{kk}\,.
\eeq

\bibliographystyle{JHEP.bst}
\bibliography{main.bib}

\end{document}